\documentclass[doublecol]{epl2}

\title{Efficient quantum key distribution scheme with pre-announcing basis}
\shorttitle{Efficient QKD pre-announcing basis} 

\author{Jingliang Gao \thanks{Email: \email{gaojl0518@gmail.com}} \and Changhua Zhu \and Heling Xiao}
\shortauthor{Jingliang Gao \etal}
\institute{State Key Laboratory of Integrated Services Networks, Xidian University, Xi'an 710071, China}

\abstract{
We devise a new quantum key distribution scheme that is more efficient than the BB84 protocol. By pre-announcing basis, Alice and Bob are more likely to use the same basis to prepare and measure the qubits, thus achieves a higher efficiency. The error analysis is revised and its security against any eavesdropping is proven briefly. Furthermore we show that, compared with the LCA scheme, our modification can be applied in more quantum channels.}

\begin{document}

\maketitle

\section{\label{sec:1}Introduction}
Quantum cryptography\cite{b.2} lies at the intersection of quantum mechanics and information theory. It utilizes fundamental theories in quantum mechanics to guarantee the information security. Quantum key distribution(QKD) is an important research direction in quantum cryptography {and} has attracted many interests recently\cite{b.1,b.2,b.3,b.4,b.5,b.6,b.7,b.8,b.9,b.10,b.11,b.12,b.13}. The best-known {QKD scheme} is the BB84 protocol\cite{b.1} {which was} suggested by Bennett and Brassard in 1984. {In their scheme, Alice and Bob randomly choose Z-basis \{$\left | 0 \right \rangle$, $\left | 1 \right \rangle$\} or X-basis \{$\left | + \right \rangle$, $\left | - \right \rangle$\} to prepare and measure the qubits. If the error rate is small enough, they perform error correction and privacy amplification {to get the final key}. Otherwise, they terminate the protocol}. This protocol has been {theoretically} proven secure\cite{b.15,b.16,b.17,b.18,b.19}. {In practice, many experimental even commercial systems based on BB84 have been proposed\cite{b.3}, most of which use photon(or weak light pulse) as qubit. An inherent defect of such systems is that photons are easy to be absorbed by the channel. For example, through 100km optical fiber, the absorption is about 20dB, which means that only one percent of the photons will reach the receiver and all the others are lost in the channel. The arrived qubits are so rare that they should not be wasted.} In BB84 scheme, Alice and Bob are using different basis with half of the probability in which case the qubits {must be} discarded. {In other word, half of the arrived qubits are wasted.} Lo, Chau and Ardehali suggested {a more efficient scheme(called LCA)\cite{b.14} in which} {the qubits are prepared in Z-basis with probability $p$ or in X-basis with probability ${\emph 1-p}$.} {By increasing ${\emph p}$, Alice and Bob are more likely to use the same basis}, so the fraction of discarded data is reduced and the efficiency is improved. To ensure the security, they employed a refined data analysis. {The error rates in Z-basis and X-basis are estimated separately}. The run is acceptable if both of the error rates, ${\emph e}_Z$ and ${\emph e}_X$, are smaller than the tolerance ${\emph e}_{max}$.

However, we note that, the {security requirement of LCA} is so stringent that the BB84 can {produce} key through some quantum channels but the LCA cannot. For instance, suppose the channel causes 15\% errors in the Z-basis, 5\% errors in the X-basis, i.e. ${\emph e}_Z=15\%$, ${\emph e}_X=5\%$. We set ${\emph e}_{max}=11\%$. {According to Shor and Preskill's proof\cite{b.19},} the BB84 protocol is secure if $\emph e_{bit},\emph e_{phase}<\emph e_{max}$, {where $\emph e_{bit}$ is the bit-flip error rate and $\emph e_{phase}$ is the phase error rate in the corresponding EPP.} It should be emphasized that $\emph e_{bit},\emph e_{phase}$ \emph {are not the same with} $\emph e_{Z},\emph e_{X}$. A confusion often happens here. In the EPP, Alice and Bob {apply Hadamard gate on half of the qubits}. For the qubits which are not Hadamard transformed, the bit-flip error rate is simply the Z-basis error rate and the phase error rate is the X-basis error rate, i.e. $\emph e_{bit}=\emph e_{Z}, \emph e_{phase}=\emph e_{X} $. Conversely, {for the case that the Hadamard gate is applied}, the bit-flip error rate is the X-basis error rate, and phase error rate is the Z-basis error rate, i.e. $\emph e_{bit}=\emph e_{X}, \emph e_{phase}=\emph e_{Z}$. The whole bit-flip and phase error rates should be the average of the two cases, i.e.
$\emph e_{bit}=\emph e_{phase}=(\emph e_{Z}+\emph e_{X})/2=10\%<11\%$. Thus for BB84, the run is acceptable and the key can be generated. For the LCA, since ${\emph e}_Z=15\%>11\%$, the run is terminated and no key is produced. From this example, we can see that the applicability of the LCA scheme is {worse than the standard BB84}.

The motivation of this paper is to explore an efficient QKD scheme which still {works when the LCA is invalid}. In the BB84 protocol, Alice and Bob choose the basis randomly, uniformly and independently. The LCA scheme removes the {uniformity to get} a gain in efficiency. Instead, our new scheme weakens the independency. {Suppose Alice encodes the qubit in the basis $b$. Before sending the qubit, she pre-announces a basis $s$. The two bases, $b$ and $s$, coincide with high probability and differ with small probability.} Bob performs measurement in the basis pre-announced, then they are more likely to be using the same basis, thus the discarded data is reduced and the efficiency is improved. We also revise the error analysis to ensure the security. The details will be explained below. {Another advantage of our modification lies in its simplicity. It doesn't need entanglement or any other auxiliaries and it can be applied in many practical QKD systems with only a little software changes. So we think it will attract some practical interests.}

{This paper is organized as following: The complete procedures of the new scheme are presented in the next section. And then the security of the scheme will be proven rigorously. At the end, the comparison between LCA and our scheme will be shown.}
\section{\label{sec:2} Description of the modification}
The explicit procedures of the new modification are presented as follows:

\noindent (1) Alice creates a random data bit string.

\noindent (2) Alice prepares another two random bit strings $\emph b$ and $\emph c$. The string $\emph b$ distributes uniformly. Alice encodes each data bit in the Z basis if $\emph b$ is 0 or X basis if $\emph b$ is 1. {The bit of the string $\emph c$ is 0 with probability $p$ or 1 with probability $\emph 1-p$}. Let $\emph s=b+c\pmod 2$, the string $\emph s$ indicates the bases that will be pre-announced.

\noindent (3) Alice publicly announces the string $\emph s$, Bob receives.

\noindent (4) Alice sends the qubits to Bob.

\noindent (5) Bob receives the qubits. He first selects a subset of them for error checking. For the checking qubits, he chooses the basis randomly and independently to perform measurement. For the other qubits, {the bases are chosen according to $\emph s$}.

\noindent (6) Bob tells Alice which qubits are selected for error checking. They share everything about these qubits.

\noindent (7) They discard the checking qubits {where} they did not use the same basis, and divide the remaining into two groups--{those for which the corresponding $\emph c$ is 0 and those for which $\emph c$ is 1.} Then {they estimate the error rate of each group, and denote them by $\emph e_{c=0}, \emph e_{c=1}$}. If both of $\emph e_{c=0}$ and $\emph e_{c=1}$ are smaller than the tolerance ${\emph e}_{max}$, they proceed to the next step, otherwise they abort the protocol.

\noindent (8) Alice announces $\emph c$ of the remaining qubits. They keep the {bits} where $\emph c=0$ as the raw key.

\noindent (9) They perform reconciliation and privacy amplification to obtain the final key.

\noindent $\emph Remark$: For the error checking qubits, the measuring basis must be chosen independently, not by $\emph s$.

The first major ingredient of our modification is pre-announcing basis. The string $\emph b$ indicates the basis that Alice encodes the qubits. {The string $\emph s$ is what she will pre-announced. The string $\emph c$ decides whether $\emph s$ coincides with $\emph b$ or not.  Alice pre-announces the string $\emph s$ before sending qubits.} Bob chooses basis according to $\emph s$, then they are using the same basis with probability $\emph p$. As $\emph p$ rises, the efficiency increases. Clearly, if $\emph p>\frac{1}{2}$, the efficiency is higher than BB84. The second major ingredient is the revised error analysis. {The qubits are divided into two subsets according to $\emph c$, and the error rates of each subset are estimated separately}. If both of the error rates are smaller than the tolerance, the run is acceptable, otherwise the protocol terminates. {It should be emphasized that for the checking qubits, Bob must choose the basis independently but not according to $s$. Otherwise, the error rates cannot be estimated correctly}. In the following section, we will show that such an error checking strategy guarantees the security of the protocol.

\section{\label{sec:3} {The proof of security}}
{Proving the security of a QKD scheme turns out to be a very tricky business\cite{b.15,b.16,b.17,b.18,b.19}. Shor and Preskill suggested a simple proof of BB84 in 2000\cite{b.19}. {They first showed that the BB84 protocol is equivalent to the entanglement-purification protocol(EPP) in the sense of security. Then since the EPP is secure, so is BB84}. In this paper, the same skill is {employed}. First, the underlying EPP protocol which is equivalent to our new modification will be given. Then we will show that the revised error analysis can guarantee the security of the EPP. Since the equivalence, our modification is secure as well.

We now give the complete procedures of underlying EPP protocol:

\noindent (1) Alice creates EPR pairs in the state $\left|\beta_{00}\right\rangle=\left(\left|00\right\rangle+\left|11\right\rangle\right)/\sqrt{2}$.

\noindent (2) Alice prepares two bit strings $\emph b$ and $\emph c$. The string $\emph b$ distributes uniformly. Alice performs a Hadamard transformation on the second qubit of EPR pair if $\emph b=1$ or does nothing if $\emph b=0$. {The string $\emph c$ is 0 with probability $\emph p$ or 1 with  $\emph 1-p$}. Let $\emph s=\emph b+\emph c\pmod{2}$, the string $\emph s$ indicates the basis that will be pre-announced.

\noindent (3) Alice publicly announces $\emph s$ and Bob receives.

\noindent (4) Alice sends the second qubit of each pair to Bob.

\noindent (5) Bob receives the qubits and publicly announces this fact.

\noindent (6) Alice announces the string $\emph c$.

\noindent (7) Bob receives c, and {acquires} the string $\emph b$ from $\emph b=\emph s+\emph c\pmod{2}$. {Then he performs Hadamard gate} on the qubits where $\emph b$ is 1.

\noindent (8) Alice and Bob publicly select a subset of EPR pairs for error checking. Then they measure their check qubits in the Z-basis and publicly share the results. They divide the checking data into two groups according to $\emph c$ and estimate the error rates $\emph e_{c=0}$, $\emph e_{c=1}$ separately. If both of $\emph e_{c=0}$ and $\emph e_{c=1}$ are smaller than $\emph e_{max}$, they proceed to the next step, otherwise they abort the protocol.

\noindent (9) Alice and Bob measure their remaining qubits according to the check matrix for a pre-determined CSS code with error tolerance $\emph e_{max}$. They share the results, compute the syndromes for the errors, and then correct their state, obtaining nearly perfect EPR pairs.

\noindent (10) Alice and Bob measure {the EPR pairs in the Z-basis to obtain the final key}.

This protocol can be reduced to a QECC protocol and then to our new scheme using the skill developed by Shor and Preskill. The details of the reductions can be found in Ref\cite{b.19,b.20,b.21}. Here we just claim its truth and prove the security of the EPP directly.

In the EPP, {the purifying procedure in step(9) will be done when the criterion $\emph e_{c=0},\emph e_{c=1}<\emph e_{max}$ satisfies. If the EPR pairs can be purified reliably, the fidelity will be very high}. By Holevo's bound\cite{b.22}, high fidelity implies low information {that can be tapped}\cite{b.17}. Therefore, to prove the security of the EPP, we just need to show that {the purification is reliable}. Furthermore, since the purification is based on the CSS code\cite{b.23,b.24}, it runs correctly only when the bit-flip and phase error rates are both smaller than the error tolerance $\emph e_{max}$. {So in summary}, we need to prove that: if $\emph e_{c=0},\emph e_{c=1}<\emph e_{max}$, then $\emph e_{bit},\emph e_{phase}<\emph e_{max}$.

{Let $\emph e_Z^{s=0}$ be the Z-basis error rate of the qubits where Alice pre-announces $s=0$, similarly we can define $\emph e_Z^{s=1}$, $\emph e_X^{s=0}$ and $\emph e_X^{s=0}$.}

In the EPP, for the qubits which are not Hadamard transformed(b=0), the bit-flip error rate is simply  the Z-basis error rate, the phase error rate is the X-basis error rate. Thus
\begin{equation}
\emph e_{bit}^{b=0}=pe_Z^{s=0}+(1-p)e_Z^{s=1}
\label{eq:1}.
\end{equation}
\begin{equation}
\emph e_{phase}^{b=0}=pe_X^{s=0}+(1-p)e_X^{s=1}
\label{eq:2}.
\end{equation}
{For the qubits where the Hadamard gates are applied(b=1)}, the bit-flip error rate is the X-basis error rate, the phase error rate is the Z-basis error rate. Thus
\begin{equation}
\emph e_{bit}^{b=1}=(1-p)e_X^{s=0}+pe_X^{s=1}
\label{eq:3}.
\end{equation}
\begin{equation}
\emph e_{phase}^{b=1}=(1-p)e_Z^{s=0}+pe_Z^{s=1}
\label{eq:4}.
\end{equation}
{Since half of the qubits are Hadamard transformed, we have}:
\begin{equation}
\emph e_{bit}=\frac{1}{2}\emph e_{bit}^{b=0}+\frac{1}{2}\emph e_{bit}^{b=1}
\label{eq:5}.
\end{equation}
\begin{equation}
\emph e_{phase}=\frac{1}{2}\emph e_{phase}^{b=0}+\frac{1}{2}\emph e_{phase}^{b=1}
\label{eq:6}.
\end{equation}
{$\emph e_{c=0}$ consists of two parts: the Z-basis error rate as s=0 and the X-basis error rate as s=1. Hence $\emph e_{c=0}$ should be the average of the two cases, i.e.}
\begin{equation}
\emph e_{c=0}=\frac{1}{2}\emph e_Z^{s=0}+\frac{1}{2}\emph e_X^{s=1}
\label{eq:7}.
\end{equation}
{$\emph e_{c=1}$ can be given by the weighted average of the X-basis error rate as s=0 and the Z-basis error rate as s=1, i.e.}
\begin{equation}
\emph e_{c=1}=\frac{1}{2}\emph e_X^{s=0}+\frac{1}{2}\emph e_Z^{s=1}
\label{eq:8}.
\end{equation}
{From Eqs. (\ref{eq:1})-(\ref{eq:8})}, we find that
\begin{equation}
e_{bit}=pe_{c=0}+(1-p)e_{c=1}
\label{eq:9}.
\end{equation}
\begin{equation}
e_{phase}=(1-p)e_{c=0}+pe_{c=1}
\label{eq:10}.
\end{equation}
Consequently, {provided $\emph e_{c=0},\emph e_{c=1}<\emph e_{max}$, we have $\emph e_{bit},\emph e_{phase}<\emph e_{max}$}, which says that, {if the error checking criterion is satisfied, the purification will be reliable. Then the EPP is secure, and so is our scheme.} This completes the proof.
}
\section{\label{sec:4} Comparison with LCA}

The comparison between LCA and our scheme is a bit confusing because they seems equivalent. However, we will show that: \emph {{Through} some quantum channels, the LCA is unable to generate key but our scheme may work}. {For example}, still set ${\emph e}_Z=15\%, {\emph e}_X=5\%, \emph e_{max}=11\%$.
As stated above, {the LCA is ineffective in such settings}. Instead, our scheme offers the opportunity. {If the error rates remain constant no matter what Alice pre-announces}, i.e. $ e_X^{s=0}=e_X^{s=1}=e_X=5\%, e_Z^{s=0}=e_Z^{s=1}=e_Z=15\%$, then from(\ref{eq:7})(\ref{eq:8}), $\emph e_{c=0}=e_{c=1}=10\%<11\%$, so the run is acceptable and the key will be generated from our scheme.

\begin{figure}
\centerline{\includegraphics{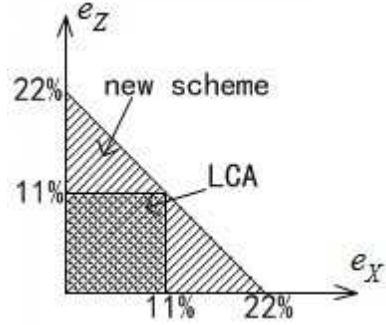}}
\caption{\label{fig:1} Comparison with LCA}
\end{figure}

More intuitively, {the error tolerances of the two schemes are shown in Fig.\ref{fig:1}.}
{Each point in the figure represents a class of quantum channels}. For example, the point(0.05, 0.03) indicates the channels with 5\% errors in X-basis and 3\% errors in Z-basis. {More points a protocol covers, more channels it can be applied in.}

For the LCA, the {security} criterion is $\emph e_Z,e_X<11\%$, which covers the dot region in Fig.\ref{fig:1} .

For our scheme, $e_{X}$ and $e_{Z}$ should be the average of the error rates as s=0 and s=1, i.e.
\begin{equation}
\emph e_{Z}=\frac{1}{2}\emph e_{Z}^{s=0}+\frac{1}{2}\emph e_{Z}^{s=1}
\label{eq:11}.
\end{equation}
\begin{equation}
\emph e_{X}=\frac{1}{2}\emph e_{X}^{s=0}+\frac{1}{2}\emph e_{X}^{s=1}
\label{eq:12}.
\end{equation}
From Eqs. (\ref{eq:7})(\ref{eq:8})(\ref{eq:11})(\ref{eq:12}),we obtain
\begin{equation}
\emph e_{Z}+\emph e_{X}=\emph e_{c=0}+\emph e_{c=1}
\label{eq:13}.
\end{equation}
{Let} $\alpha=\emph e_{c=1}/\emph e_{c=0}$, then the security criterion $\emph e_{c=0},\emph e_{c=1}<\emph e_{max}$ becomes
\begin{equation}
e_{Z}+e_{X} < \min\left\{(1+\alpha)\emph e_{max}, (1+\frac{1}{\alpha})\emph e_{max}\right\}
\label{eq:14}
\end{equation}
{When $\alpha=1$, it covers the maximum area, i.e.}
\begin{equation}
\frac{1}{2}(e_{X}+e_{Z})<e_{max}
\label{eq:15}.
\end{equation}
{which is shown as the slashed region in Fig.\ref{fig:1}}.
Clearly, the coverage of the new scheme is larger than that of LCA.

{It should be clarified that, whether the new scheme generates key or not, not only depends on $e_{X}$ and $e_{Z}$, but also depends on the parameter $\alpha$. From(\ref{eq:14}), if $\alpha$ takes other value, e.g. $\alpha=2$, the criterion becomes $e_X+e_Z <\frac{3}{2}e_{max}$, which covers a smaller area than shown in Fig.\ref{fig:1}. In this case, the new scheme cannot produce key when $e_{X}=0.05,e_{Z}=0.15$ because $e_X+e_Z = 0.2>\frac{3}{2}e_{max}$. So the new scheme is not always effective when LCA is invalid. However, it does provide the possibility. Thus our scheme may be a better choice in practice.\\
The parameter $\alpha$ can be used to estimate how much pre-announced information was utilized by Eve. If there is no Eve, $\alpha$ should be 1. If Eve exists and the pre-announced basis is used for eavesdropping, more errors will occur in the part of the qubits where the pre-announced bases are incorrect(where c=1). Then $\alpha$ will be larger than 1. By observing $\alpha$, Alice and Bob can get some knowledge about Eve's intercepting strategy.}\\
{In LCA's paper, they declared that their scheme can be further improved by combining it with two-way classical communication. Luckily, our scheme also benefits from the same extension. It was shown in\cite{b.25} that the $e_{max}$ can be increased up to 18.9\% if the two-way classical communication is applied. Combining with this result, the best coverage (when $\alpha=1$) of our scheme is expanded to $e_{X}+e_{Z}<37.8\%$.}

\section{\label{sec:5} Conclusion}
In this paper, we suggest a new QKD scheme. By pre-announcing a batch of bases, Alice and Bob use the same basis with high probability, thus achieve a high efficiency. We revise the error checking criterion and prove its security. We also compared our scheme with the LCA scheme and show that our scheme may be a better choice in some quantum channels.




\acknowledgments
This work is supported by the National Natural Science Foundation of China Grant No.61271174.

\end{document}